\providecommand*{\acmlayout}{1}
\providecommand*{\lncslayout}{2}
\newcommand*{\chosenlayout}{\acmlayout}
\providecommand*{\acmlayout}{1}
\providecommand*{\lncslayout}{2}
\providecommand*{\chosenlayout}{\acmlayout}

\ifnum\chosenlayout=\acmlayout\relax
  \documentclass[sigconf]{acmart}
\fi
\ifnum\chosenlayout=\lncslayout\relax
  \documentclass[runningheads]{llncs}
\fi

\usepackage[utf8]{inputenc}
\usepackage[T1]{fontenc}
\usepackage{xspace}
\usepackage{xcolor}
\usepackage{ifthen}
\usepackage{pifont}
\usepackage{enumitem}
\usepackage{listings}
\usepackage{lstlinebgrd}
\usepackage{suffix}
\usepackage{comment}
\usepackage{algorithm}
\usepackage{algpseudocode}
\usepackage{hyperref}
\usepackage{todonotes}
\usepackage[nohypertypes={main,acronym}]{glossaries-extra}
\usepackage{mathtools}
\usepackage{siunitx} 
\usepackage{interval} 
\usepackage{relsize} 
\ifnum\chosenlayout=\lncslayout\relax
  \usepackage{orcidlink} 
  \usepackage{wrapfig} 
  \usepackage{caption} 
\fi

\newboolean{our_draft}
\setboolean{our_draft}{false}
\newboolean{colored}
\setboolean{colored}{false}
\newboolean{doublecolumn}
\ifnum\chosenlayout=\acmlayout\relax
  \setboolean{doublecolumn}{true}
\fi
\ifnum\chosenlayout=\lncslayout\relax
  \setboolean{doublecolumn}{false}
\fi

\newcommand{\Invariant}[1]{\Statex \textbf{invariant} #1}
\newcommand{\Ternary}[3]{\textbf{if } #1 \textbf{ then } #2 \textbf{ else } #3 \textbf{ end if}}
\DeclareMathOperator{\isodd}{isodd}
\definecolor{backgroundhighlight}{HTML}{7FADDC}
\newcommand{\mhighlight}[1]{%
  \mkern2mu
  \colorbox{backgroundhighlight}{$\displaystyle\mkern-8mu#1\mkern-4mu$}%
}

\newcommand{\ie}{i.e.,\xspace}
\newcommand{\eg}{e.g.,\xspace}

\setglossarystyle{list} 
\setabbreviationstyle[acronym]{long-short-sc}

\newglossary[glignoredl]{ignored}{glignored}{glignoredin}{Ignored Glossary}

\definecolor{commentcolor}{HTML}{747678}
\definecolor{ghostkeywordcolor}{HTML}{CC5500}
\definecolor{ghostcodebgcolor}{HTML}{e1e2e2}
\newcommand*{\ghostcodebgcolor}{gray}

\makeatletter
\newcommand{\highlightlines}[2][lightblue]{%
  \def\@tempcolor{#1}%
  \processranges{#2}%
}
\ExplSyntaxOn
\newcommand{\processranges}[1]{%
  \seq_set_split:Nnn \l_tmpa_seq {,} {#1}%
  \seq_map_inline:Nn \l_tmpa_seq {%
    \@checkrange{##1}%
  }%
}
\ExplSyntaxOff
\newcommand{\@checkrange}[1]{%
  \expandafter\@splitrange#1-\relax%
}
\def\@splitrange#1-#2\relax{%
  \def\@first{#1}%
  \def\@second{#2}%
  \ifx\@second\empty%
    \ifnum\value{lstnumber}=\@first%
      \color{\@tempcolor}%
    \fi%
  \else%
    \edef\@second{\expandafter\@gobble@dash\@second}%
    \ifnum\value{lstnumber}>\numexpr\@first-1\relax%
      \ifnum\value{lstnumber}<\numexpr\@second+1\relax%
        \color{\@tempcolor}%
      \fi%
    \fi%
  \fi%
}
\def\@gobble@dash#1-{#1}
\makeatother

\lstdefinestyle{figurelistingstyle}{
  floatplacement={tbp},
  captionpos=b,
  frame=lines,
  numbers=left,
  numberblanklines=false,
  xleftmargin=8pt,
  xrightmargin=8pt,
  numberstyle=\scriptsize,
  breaklines=true,
  basicstyle={\ttfamily\ifthenelse{\boolean{doublecolumn}}{\footnotesize}{\scriptsize}},
  commentstyle={\color{commentcolor}\textit},
  keywordstyle={[1]\color[HTML]{0005FF}},
  keywordstyle={[2]\color{ghostkeywordcolor}},
  keywordstyle={[3]\color[HTML]{EC008C}},
}

\lstdefinelanguage{gobra}{
  language=go,
  sensitive=true,
  morecomment=[l]{//},
  morecomment=[s]{/*}{*/},
  moredelim=*[l]{//@},
  moredelim=*[s]{/*@}{@*/},
  morekeywords=[1]{ 
    pred, implements, set, match, foreach, of, integer
  },
  morekeywords=[2]{ 
    requires, ensures, preserves, invariant, pure, unfolding, in, forall, acc, let, ghost, trusted, decreases, opaque
  },
  morekeywords=[3]{ 
    fold, unfold,
    assume, assert, inhale, exhale, reveal
  },
  style=figurelistingstyle,
  mathescape=true,
  moredelim=**[is][\normalfont\itshape]{'}{'}
}

\lstnewenvironment{gobracode}[1][]{
  \lstset{
    language=gobra,
    #1
  }
}{}

\lstnewenvironment{gobracodeinminipage}[1][]{
  \lstset{
    language=gobra,
    xleftmargin=0pt,
    xrightmargin=16pt,
    #1
  }
}{}

\newcommand{\code}[1]{\mbox{\lstinline[language=gobra,basicstyle=\ttfamily]~#1~}}
\newcommand{\codebw}[1]{\mbox{\lstinline[language=gobra,basicstyle=\ttfamily\color{black},keywordstyle={[1]\color{black}},keywordstyle={[2]\color{black}},keywordstyle={[3]\color{black}}]~#1~}}

\lstdefinelanguage{tamarin}{
  sensitive=true,
  morecomment=[l]{//},
  morecomment=[s]{/*}{*/},
  morekeywords=[1]{
    aenc, sdec, senc, sdec, sign, verify, hashing, multiset, revealSign, revealVerify, getMessage, true
  },
  morekeywords=[2]{
    axiom, begin, builtins, end, equations, functions, heuristic, in, let, options, predicate, predicates, property, protocol, restriction, section, subsection, text, theory, verdictfunction
  },
  morekeywords=[3]{
    new, in, out, lookup, as, in, else, if, lock, unlock, event, insert, delete, then, account, accounts, for, parties, otherwise
  },
  morekeywords=[4]{
    use_induction, sources, reuse, hide_lemma, left, right
  },
  morekeywords=[5]{
    F, T, All, Ex
  },
  style=figurelistingstyle,
  mathescape=true
}

\lstnewenvironment{tamarincode}[1][]{
  \lstset{
    language=tamarin,
    #1
  }
}{}

\lstnewenvironment{tamarincodeinminipage}[1][]{
  \lstset{
    language=tamarin,
    xleftmargin=0pt,
    xrightmargin=16pt,
    #1
  }
}{}

\lstdefinelanguage{fstar}{
  sensitive=true,
  morecomment=[l]{//},
  morecomment=[s]{/*}{*/},
  morekeywords=[1]{ 
    module, open, val, let, in, assert, fun
  },
  style=figurelistingstyle,
  mathescape=true,
  moredelim=**[is][\normalfont\itshape]{'}{'}
}

\lstnewenvironment{fstarcode}[1][]{
  \lstset{
    language=fstar,
    #1
  }
}{}

\lstnewenvironment{fstarcodeinminipage}[1][]{
  \lstset{
    language=fstar,
    xleftmargin=0pt,
    xrightmargin=16pt,
    #1
  }
}{}

\newcommand{\mypar}[1]{\myparbold{#1}}
\newcommand{\myparbold}[1]{\smallskip\noindent\textbf{#1.}}

\ifthenelse{\boolean{our_draft}}{
  \newcommand{\fix}[1]{\sidenote{\textcolor{orange}{#1}}}
}{
  \newcommand{\fix}[1]{}
}

\ifthenelse{\boolean{our_draft}}{
  \ifthenelse{\boolean{colored}}{ 
    
    \newenvironment{newtext}{\color{blue}}{}
    \newcommand{\remove}[1]{\textcolor{red}{#1}}

  }{
    
    \newenvironment{newtext}{}{}
    \newcommand{\remove}[1]{}
    \excludecomment{removetext}
    
  }
  \newcommand{\drafttodo}[1]{\todo{#1}}
}{
  \ifthenelse{\boolean{colored}}{ 

  }{
  }
  \newcommand{\drafttodo}[1]{}
}

\definecolor{lightblue}{rgb}{.651,.745,0.875}

\usepackage{newfloat}	
\DeclareFloatingEnvironment[
    listname={List of Protocols},
    name=Protocol,
    placement=!ht,
]{protocol}

\DeclareFloatingEnvironment[
    listname={List of Attacks},
    name=Attack,
    placement=!ht,
    within=section,
]{attack}

\newcommand*{\proofsketchname}{Proof sketch}

\renewcommand*{\ie}{i.e.,\xspace}

\renewcommand*{\eg}{e.g.,\xspace}

\newcommand*{\cf}{cf.\xspace}

\newcommand*{\wrt}{w.r.t.\xspace}
\newcommand{\quotes}[1]{``#1''}

\newcommand*{\Figref}[1]{Fig.~\ref{fig:#1}}
\newcommand*{\figref}{\Figref}
\newcommand*{\Tabref}[1]{Tab.~\ref{tab:#1}}
\newcommand*{\tabref}{\Tabref}

\newcommand*{\Secref}[1]{Sec.~\ref{sec:#1}}
\newcommand*{\secref}{\Secref}

\newcommand*{\Lineref}[1]{Line~\ref{line:#1}}
\newcommand*{\lineref}[1]{line~\ref{line:#1}}
\newcommand*{\Linerange}[2]{Lines~\ref{line:#1}--\ref{line:#2}}
\newcommand*{\linerange}[2]{lines~\ref{line:#1}--\ref{line:#2}}

\newcommand*{\Exampleref}[1]{Example~\ref{example:#1}}
\newcommand*{\exampleref}{\Exampleref}
\newcommand*{\Algref}[1]{Alg.~\ref{alg:#1}}
\renewcommand*{\algref}{\Algref}
\WithSuffix{\newcommand*}\Figref*[1]{Fig.~#1}
\WithSuffix{\newcommand*}\figref*{\Figref*}
\WithSuffix{\newcommand*}\Secref*[1]{Sec.~#1}
\WithSuffix{\newcommand*}\secref*{\Secref*}
\WithSuffix{\newcommand*}\Lineref*[1]{Line~#1}
\WithSuffix{\newcommand*}\lineref*[1]{line~#1}
\WithSuffix{\newcommand*}\Lines*[1]{Lines~{#1}}
\WithSuffix{\newcommand*}\lines*[1]{lines~{#1}}
\WithSuffix{\newcommand*}\Linerange*[1]{Lines~{#1}}
\WithSuffix{\newcommand*}\linerange*[1]{lines~{#1}}
\WithSuffix{\newcommand*}\Exampleref*[1]{Example~#1}
\WithSuffix{\newcommand*}\exampleref*{\Exampleref*}

\newcommand{\lbbar}{\{\kern-0.5ex|} 
\newcommand{\rbbar}{|\kern-0.5ex\}} 

\newcommand*{\tool}[1]{\textsc{#1}\xspace}

\newcommand*{\gobra}{\tool{Gobra}}
\newcommand*{\viper}{\tool{Viper}}
\newcommand*{\lean}{\tool{Lean}}
\newcommand*{\zthree}{\tool{Z3}}

\newcommand*{\dafny}{\tool{Dafny}}

\newcommand*{\fstar}{\tool{F$^{\star}$}}

\newcommand*{\haclstar}{\tool{HACL$^{\star}$}}
\newcommand*{\vale}{\tool{Vale}}

\newcommand*{\libcrux}{\tool{libcrux}}

\newcommand*{\rocq}{\tool{Rocq}}

\newcommand*{\easycrypt}{\tool{EasyCrypt}}

\newcommand*{\evercrypt}{\tool{EverCrypt}}

\newcommand*{\jasmin}{\tool{Jasmin}}
\newcommand*{\libjbn}{\tool{libjbn}}

\newcommand*{\fiatcrypto}{\tool{Fiat Cryptography}}

\newcommand*{\boringssl}{\tool{BoringSSL}}
\newcommand*{\openssl}{\tool{OpenSSL}}

\newcommand*{\goextgcd}{\code{extendedGCD}\xspace}
\newcommand*{\gomodinv}{\code{InverseVarTime}\xspace}
\newcommand*{\gogcd}{\code{GCDVarTime}\xspace}

\newacronym[
    shortplural={LOC},
    longplural={lines of code}
]{LOC}{LOC}{line of code}
\newacronym[
    shortplural={LOS},
    longplural={lines of specification}
]{LOS}{LOS}{line of specification}
\newacronym[
    shortplural={ALOC},
    longplural={auxiliary lines of code}
]{ALOC}{ALOC}{auxiliary line of code}
\newacronym{FIPS}{FIPS}{Federal Information Processing Standards}
\newacronym{GCD}{GCD}{Greatest Common Divisor}

\newcommand*{\ac}[1]{\gls{#1}}
\newcommand*{\acp}[1]{\glspl{#1}} 
\newcommand*{\acsp}[1]{\glsfmtshortpl{#1}} 
\newcommand*{\acptitle}[1]{\glsfmtfullpl{#1}} 

\ifnum\chosenlayout=\acmlayout\relax
\newcommand*{\keywordsep}{, }
\fi
\ifnum\chosenlayout=\lncslayout\relax
\newcommand*{\keywordsep}{ \and}
\fi
\newcommand*{\mykeywords}{extended euclidean algorithm\keywordsep go language\keywordsep standard library\keywordsep separation logic\keywordsep automated verification\keywordsep formal methods.}

\ifnum\chosenlayout=\acmlayout

  \keywords{\mykeywords}

  \acmYear{2026}
  \acmConference[FCS '26]{2026 Workshop on Foundations of Computer Security}{July 25, 2026}{Lisbon, Portugal}
  \acmBooktitle{2026 Workshop on Foundations of Computer Security (FCS '26), July 25, 2026, Lisbon, Portugal}

  \copyrightyear{}
  \acmISBN{}
  \acmDOI{}
\fi

\begin{document}

\ifnum\chosenlayout=\acmlayout\relax
  \settopmatter{printfolios=true, printacmref=false} 
  \setcopyright{none} 
\fi

\title{GCD: Garbled, Corrected, Demonstrandum}
\subtitle{Fixing and Proving Go's Extended GCD Implementation}

\ifnum\chosenlayout=\acmlayout\relax
  \author{Linard Arquint}
  \orcid{0000-0002-6230-8014}
  \affiliation{%
    \institution{National University of Singapore}
    \country{Singapore}
  }
\fi
\ifnum\chosenlayout=\lncslayout\relax
  \author{Linard Arquint \orcidlink{0000-0002-6230-8014}}
  \authorrunning{Linard Arquint}
  \institute{National University of Singapore}
\fi

\ifnum\chosenlayout=\acmlayout\relax
  \begin{abstract}
We verify the \goextgcd implementation in Go's standard library (\texttt{crypto\slash{}internal\slash{}fips140\slash{}bigmod}), which plays a crucial role in the generation of RSA key pairs.
Even though the Go implementation is supposedly a direct port of \boringssl's implementation, we uncovered two~deviations that each break the invariants of an existing proof for \boringssl:
(1)~the Go implementation deviates in the way coefficients are updated, and (2)~it permits a larger input domain.
We prove correctness and termination of both the existing Go implementation and a fixed implementation using \gobra, a deductive program verifier for Go.
Where necessary, we used \lean to prove key lemmata on non-linear arithmetic, which we import into \gobra.
For the existing implementation, we devise new invariants that accommodate the deviating coefficient updates;
for the fixed implementation, we align the coefficient updates with \boringssl's and port the existing proof.
In both cases, we extend the proof to cover the larger input domain.
Our fixed implementation updates coefficients in-place and, thus, reduces memory allocations, memory initializations, and memory copying operations resulting in an average speedup of \qty{24}{\percent} over the existing implementation.

Our verification effort reveals three~key insights: subtle bugs can slip into even well-reviewed code with surprising ease; formal verification is a powerful tool for uncovering them; and AI agents can facilitate the verification process by iteratively refining invariants and lemmata based on \gobra's error messages.

\ifnum\chosenlayout=\lncslayout\relax
\keywords{\mykeywords}
\fi
\end{abstract}

\fi

\maketitle

\ifnum\chosenlayout=\lncslayout\relax
  
\fi

\section{Introduction}
\label{sec:introduction}
\looseness=-1
Standard libraries are a critical part of the software ecosystem, as they provide fundamental building blocks for applications and are, thus, widely used and heavily scrutinized.
The Go programming language provides numerous cryptographic functionality in its standard library, ranging from random number generation and RSA to TLS and SSH.
Go's standard library implements algorithms approved by NIST's \ac{FIPS} 140-3~\cite{fips140-3}, which states security requirements for applications handling sensitive data in the US~government and regulated industries such as finance and healthcare.

Correctness of a standard library and in particular its cryptographic code is crucial due to their widespread use and the security implications of bugs.
Program verification enables us to rigorously reason about the correctness of an implementation, taking all possible inputs, execution paths, and memory states into account, thus uncovering bugs that may be missed by testing.

In this work, we focus on the implementation of the Extended Euclidean Algorithm, extended \ac{GCD} for short, in Go's standard library.
This implementation is crucial for RSA key generation and thus all applications that depend on RSA, including TLS, SSH, and PGP, as it computes the modular inverse of an RSA public exponent.
This modular inverse forms the core of any RSA private key, which is regulated by \ac{FIPS} 140-3.

The currently used extended \ac{GCD} implementation in Go's standard library was introduced in Go~1.24 in an effort to natively support \ac{FIPS} 140-3 approved algorithms.
Previous versions of Go relied on \boringssl, Google's fork of \openssl, for \ac{FIPS}-compliance, which however was limited to just a few supported target platforms.

Our results show that the justification for the correctness of the currently used extended \ac{GCD} implementation in Go's standard library is flawed.
The implementation explicitly appeals to \boringssl's correctness proof~\cite{fiatcrypto-extended-gcd-1,fiatcrypto-extended-gcd-2} in the \rocq proof assistant~\cite{rocq-prover} because the implementation claims to be a direct port of \boringssl's implementation.
However, we found that the implementation deviates from \boringssl's implementation in two~critical ways, both breaking the invariants on which the proof relies.
Despite code reviews by three~reviewers~\cite{go-extended-gcd}, these deviations were not caught, which demonstrates how easily subtle bugs can be introduced even in well-reviewed code.

\mypar{Contributions}
We identify deviations of Go's extended \ac{GCD} implementation \wrt \boringssl's implementation and propose a fixed implementation.
Since the fixed implementation avoids costly memory operations by performing in-place updates, it achieves an average speedup of \qty{24}{\percent} over the existing implementation.
We prove correctness and termination of both the existing and the fixed implementation using \gobra~\cite{DBLP:conf/cav/WolfACOPM21}, a deductive program verifier for Go, and use \lean to prove key lemmata that are out of scope for \gobra.
We port \boringssl's correctness proof from \rocq to \gobra, extend it to cover the larger input domain that the Go implementation permits, and adapt it to the existing implementation's deviations.
By performing the proof directly on the Go implementation, we close the gap between the implementation and its proof, which allowed the discovered deviations to stay uncovered until now.
Since \gobra takes only about \qty{34}{\second} to prove both implementations' correctness, we integrate \gobra into continuous integration to automatically check correctness on every change.
This integration ensures that we detect future deviations, which is not the case for \boringssl's proof.

\mypar{Disclosure}
\looseness=-1
We disclosed the deviations and proposed fixes to one of the primary code owners of the Go cryptography standard library on March 5, 2026.
Since we did not yet have a correctness proof for the existing implementation, we confirmed together that the deviations result in at most an availability issue for RSA key generation.
In particular, an incorrectly computed modular inverse is caught by defensive checks in the RSA key generation code, which would then return an error instead of an incorrect key.
By now, the proposed fixes have been merged into Go's master branch and are scheduled for inclusion in Go~1.28~\cite{gcd-fix-pr}.

\section{Extended \ac{GCD}}
\label{sec:extendedgcd}
The extended \ac{GCD} algorithm computes, given integers $a$ and $n$, the greatest common divisor $\gcd(a, n)$ together with the B\'ezout coefficients $A$ and $B$ satisfying $\gcd(a, n) = A * a - B * n$.
When $\gcd(a, n) = 1$, the coefficient $A$ is the modular inverse of $a$ modulo $n$, \ie $A * a \equiv 1 \pmod{n}$.
This operation is fundamental in cryptography for RSA key generation as well as elliptic curve cryptography, which relies on modular inverses in prime fields.

RSA key generation uses the extended \ac{GCD} algorithm in the following way.
First, two~large, secret prime numbers~$p$ and~$q$ are generated, and the public modulus is computed as $N = p * q$.
Next, the key generation computes the secret exponent of the multiplicative group of integers modulo $N$, which is $\lambda(N) = \mathrm{lcm}(p-1, q-1) = \frac{(p-1)*(q-1)}{\gcd(p-1, q-1)}$.
Then, a public exponent~$e$ is chosen such that $1 < e < \lambda(N)$ and $\gcd(e, \lambda(N)) = 1$.
Finally, the private key exponent~$d$ is computed as the modular inverse of $e$ modulo $\lambda(N)$, \ie $d \equiv e^{-1} \pmod {\lambda(N)}$.
Thanks to the extended \ac{GCD} algorithm, we can compute $d$ efficiently because $d$ corresponds to the B\'ezout coefficient~$A$ for $e$ and $\lambda(N)$.

\mypar{Extended \ac{GCD} in \fiatcrypto}
\looseness=-1
\fiatcrypto~\cite{DBLP:conf/sp/ErbsenPGSC19} is a framework for the \rocq proof assistant that enables expressing and proving cryptographic arithmetic in a high-level language and extracts efficient, architecture-specific, and formally verified C~implementations.
While \fiatcrypto has been used to prove the correctness of extended \ac{GCD} as implemented in \boringssl~\cite{fiatcrypto-extended-gcd-1,fiatcrypto-extended-gcd-2}, the actual implementation used in \boringssl is handwritten, \ie not extracted by \fiatcrypto.
Hence, there is no formal link between \boringssl's implementation and its proof, so regressions in the implementation may go unnoticed unless the proof is updated accordingly.

The extended \ac{GCD} algorithm used by \boringssl loosely follows Algorithm~14.61 from the Handbook of Applied Cryptography~\cite{DBLP:books/crc/MenezesOV96}, but merges loops, rearranges steps, and uses constant-time operations~\cite{fiatcrypto-extended-gcd-1,fiatcrypto-extended-gcd-2}.
We provide this algorithm in \algref{extgcd-fiatcrypto} but rewrite constant-time operations into variable-time operations for simplicity.
Although the algorithm computes and stores the \ac{GCD} in variable~$u$, it does not return this value; instead, the algorithm returns the modular inverse if the \ac{GCD} is~1 and an error otherwise.
In each loop iteration, the algorithm halves either $u$ or $v$ and maintains the following invariants.
\begin{align}
u &= A * a - B * n \label{eq:inv-u} \\
v &= D * n - C * a \label{eq:inv-v} \\
\gcd(u, v) &= \gcd(a, n) \label{eq:inv-gcd} \\
\left(\isodd(u) \lor \isodd(v)\right) &\land \left(\isodd(a) \lor \isodd(n)\right) \label{eq:inv-odd}
\end{align}

\ifthenelse{\boolean{doublecolumn}}{%
\begin{algorithm}}{%
\begin{wrapfigure}{R}{0.55\textwidth}
\begin{minipage}{0.52\textwidth}
\captionsetup{type=algorithm,position=top}
\algrenewcommand\alglinenumber[1]{\scriptsize#1}
\scriptsize
}
\caption{Binary Extended GCD in \boringssl and \fiatcrypto. For simplicity, we rewrite constant-time as variable-time operations.}
\label{alg:extgcd-fiatcrypto}
\begin{algorithmic}[1]
\Require Two positive integers $a$ and $n$, such that $a \neq 0$, $n \neq 0$, $a < n$, and not both $a$ and $n$ are even. $\mathrm{abits}$ and $\mathrm{nbits}$ are the bit lengths of $a$ and $n$, respectively.
\Ensure Error or integer $A$, such that $(A * a) \equiv 1 \pmod{n}$.
\State $u \gets a$, \quad $v \gets n$, \quad $i \gets 0$
\State $A \gets 1$, \quad $B \gets 0$, \quad $C \gets 0$, \quad $D \gets 1$

\While{$i < \mathrm{abits} + \mathrm{nbits}$}
    \Invariant{$\eqref{eq:inv-u} \land \eqref{eq:inv-v} \land \eqref{eq:inv-gcd} \land \eqref{eq:inv-odd}$}
    \If{$u$ and $v$ are both odd} \label{line:both-odd-start}
        \If{$v < u$}
            \State $u \gets u - v$
            \If{$A + C < n$} \label{line:coef-AB-update-start}
                \State $A \gets A + C$, \quad $B \gets B + D$
            \Else
                \State $A \gets A + C - n$, \quad $B \gets B + D - a$
            \EndIf \label{line:coef-AB-update-end}
        \Else
            \State $v \gets v - u$
            \If{$A + C < n$} \label{line:coef-CD-update-start}
                \State $C \gets A + C$, \quad $D \gets B + D$
            \Else
                \State $C \gets A + C - n$, \quad $D \gets B + D - a$
            \EndIf \label{line:coef-CD-update-end}
        \EndIf
    \EndIf \label{line:both-odd-end}
    \If{$u$ is even} \label{line:one-even-start}
        \State $u \gets u/2$
        \If{$A$ or $B$ is odd}
            \State $A \gets (A+n)/2$, \quad $B \gets (B+a)/2$
        \Else
            \State $A \gets A/2$, \quad $B \gets B/2$
        \EndIf
    \Else
        \State $v \gets v/2$
        \If{$C$ or $D$ is odd}
            \State $C \gets (C+n)/2$, \quad $D \gets (D+a)/2$
        \Else
            \State $C \gets C/2$, \quad $D \gets D/2$
        \EndIf
    \EndIf \label{line:one-even-end}
    \State $i \gets i + 1$
\EndWhile
\State \Return \Ternary{$u = 1$}{$A$}{\textbf{error}}
\end{algorithmic}
\ifthenelse{\boolean{doublecolumn}}{%
\end{algorithm}}{%
\vspace{-2.5\baselineskip}
\end{minipage} 
\end{wrapfigure}}

\noindent
with bounds $0 < n < 2^\mathrm{nbits}$, $0 < a < 2^\mathrm{abits}$, $a < n$, $0 < u \leq a$, $0 \leq v \leq n$, $0 \leq A < n$, $0 \leq B < a$, $0 \leq C < n$, $0 \leq D \leq a$.
$\isodd(x)$ denotes whether $x$ is odd, \ie $x \equiv 1 \pmod{2}$.

Each loop iteration consists of two~parts:
(1)~if both $u$ and $v$ are odd, the algorithm subtracts the smaller from the larger, which ensures that exactly one of them becomes even, and updates the coefficients $A$, $B$, $C$, and $D$ accordingly~(\linerange{both-odd-start}{both-odd-end});
(2)~halve either $u$ or $v$ (depending on which is even) and update the coefficients accordingly to maintain the invariants~(\linerange{one-even-start}{one-even-end}).
Between these two~parts, \ie on \lineref{one-even-start}, exactly one of $u$ and $v$ is even and the invariants hold.

\mypar{Extended \ac{GCD} in Go}
The Go standard library implements the extended \ac{GCD} algorithm in the function \goextgcd in the \texttt{crypto\slash{}internal\slash{}fips140\slash{}bigmod} package~\cite{extended-gcd-flawed}.
As stated in the description of the corresponding change~\cite{go-extended-gcd}, the Go implementation is supposedly a direct port of \boringssl's implementation, and code comments point to \fiatcrypto's proof as justification for the Go implementation's correctness.

The \goextgcd function returns not only the B\'ezout coefficient~$A$ (like \boringssl) but also the \ac{GCD} value itself.
While \goextgcd is a package-private function, the standard library exposes these return values through two~different functions to clients.
\gogcd returns the \ac{GCD} value, while \gomodinv returns the modular inverse if the \ac{GCD} is~1, and an error otherwise.

The implementation's correctness is crucial as RSA key generation within Go's standard library uses both~functions.
\gogcd is used to compute $\lambda(N)$, while \gomodinv is used to compute the private exponent~$d$ as the modular inverse of the public exponent~$e$ modulo $\lambda(N)$.

Correctness of the implementation is claimed~\cite{go-extended-gcd} only by code comments pointing to the \fiatcrypto proof.
In particular, a comment lists the very same algorithmic changes as also found in the \fiatcrypto proof documenting \boringssl's changes \wrt the original Algorithm~14.61 from the Handbook of Applied Cryptography.
No comment indicates that the Go implementation performs any additional algorithmic changes.
However, we found that the Go implementation deviates substantially from the \boringssl and \fiatcrypto implementation, which is critical as the code comments imply that the correctness of the Go implementation directly follows from \fiatcrypto's proof.

We found two~classes of deviations in the Go implementation \wrt the \boringssl and \fiatcrypto implementation.
The first~class consists of two~major but subtle deviations;
major as each such deviation breaks the invariants on which \fiatcrypto's proof relies, and subtle as neither the code's author nor three~code reviewers spotted them.
We found these two~deviations only after a failed proof attempt and discuss them in detail in \secref{major-deviations}.

The second~class consists of three~minor deviations.
First, the Go implementation uses variable-time operations and, thus, features more complex control flow, making a statement-by-statement correspondence between the two~implementations non-trivial.
Second, the Go implementation loops until $v$ becomes zero, while \boringssl's and \fiatcrypto's implementations use a fixed number of iterations, which corresponds to the sum of the bit lengths of the inputs.
This impacts the termination argument.
Third, the Go implementation of \goextgcd does not return an error when the \ac{GCD} value is different from~1, while \boringssl does.
We prove correctness of these minor deviations by adapting \fiatcrypto's proof in a straightforward way.
We report on the full proof effort in \secref{verified-impl}.

\section{Major Deviations}
\label{sec:major-deviations}
We report and address the two~major deviations that we found while carrying out the proofs on the Go implementation.
While they are subtle and, thus, have slipped through three~code reviews, both deviations are critical as each one breaks the invariants on which \fiatcrypto's proof relies.

\subsection{Unsynchronized Subtractions Deviation}
\label{sec:wrapping-bug}
The first~major deviation that we found in the Go implementation concerns the updating of the coefficients $A$, $B$, $C$, and $D$ in the first~part of every loop iteration, namely \linerange{coef-AB-update-start}{coef-AB-update-end} and \linerange{coef-CD-update-start}{coef-CD-update-end} in \algref{extgcd-fiatcrypto}.
\boringssl and \fiatcrypto consider whether the sum $A + C$ exceeds the modulus~$n$ to decide whether to reduce $A + C$ \emph{and} $B + D$ by subtracting $n$ and $a$, respectively, thus, ensuring that both reductions happen synchronously.
Synchronizing these reductions is crucial for maintaining the loop invariants~\eqref{eq:inv-u} and \eqref{eq:inv-v}.
We illustrate this by focusing on the case where $u$ and $v$ are both odd and $v \geq u$ (the case where $v < u$ is analogous).
At the beginning of a loop iteration, the invariants hold, and the updates to $v$, $C$, and $D$ maintain the invariants as follows.
We use primed variables to denote the updated values.
Ignore the summands highlighted in blue if $A + C < n$.
If $A + C \geq n$, the algorithm performs a synchronized reduction using the highlighted summands, which cancel out.
\begin{align*}
    v' &= v - u\\
       &= (D*n - C*a) - (A*a - B*n)\\
       &= B*n + D*n - A*a - C*a \mhighlight{{}- a*n + a*n}\\
       &= (B+D\mhighlight{{}-a})*n - (A+C\mhighlight{{}-n})*a\\
       &= D'*n - C'*a
\end{align*}

The Go implementation, however, performs unsynchronized reductions by updating the coefficients as
\begin{algorithmic}
\State $A \gets \Ternary{A + C < n}{A + C}{A + C - n}$
\State $B \gets \Ternary{B + D < a}{B + D}{B + D - a}$
\end{algorithmic}
and
\begin{algorithmic}
\State $C \gets \Ternary{A + C < n}{A + C}{A + C - n}$
\State $D \gets \Ternary{B + D < a}{B + D}{B + D - a}$
\end{algorithmic}
instead of \linerange{coef-AB-update-start}{coef-AB-update-end} and \linerange{coef-CD-update-start}{coef-CD-update-end}, respectively.
Whenever only one of the two~sums gets reduced, one of the two~summands highlighted in blue remains without a corresponding canceling summand, which breaks invariant~\eqref{eq:inv-v} (or \eqref{eq:inv-u} in the case $v < u$).
Nevertheless, we verify both the existing implementation, for which we adapt the invariants as described next, and a fixed implementation that resolves this deviation.
The code's author confirmed the deviation to be unintentional, and our fix additionally improves runtime performance~(\cf \secref{verified-impl-eval}).

\mypar{Verifying the Existing Implementation}
Although unsynchronized reductions do occur in the existing implementation (\eg for $a = 1$ and $n = 3$), they do not affect the result.
For $a < n$, the invariants~\eqref{eq:inv-u} and \eqref{eq:inv-v} together with bounds imply that $B + D \geq a$ if $A + C \geq n$, and $B + D \leq a$ otherwise~\cite{fiatcrypto-extended-gcd-1,fiatcrypto-extended-gcd-2}.
Hence, the reductions are unsynchronized only if $A + C < n$ and $B + D = a$, in which case the invariants imply $v - u = (n - A - C)*a > 0$.
Thus, only the update of $C$ and $D$ can be affected, which breaks invariant~\eqref{eq:inv-v}, while invariant~\eqref{eq:inv-u} continues to hold.
In this case, $u = \gcd(a, n)$ already holds.
To prove this, we additionally track that $a$ is a non-negative integer linear combination of $u$ and $v$, \ie $a = P*u + Q*v$ for some integers $P, Q \geq 0$.
Since $v - u \geq a$, $v > a$ holds and, thus, $Q$ must be~$0$.
Therefore, $u$ divides $a$ and, because of $v = u + (n - A - C)*a$, also $v$.
Hence, $u = \gcd(u, v) = \gcd(a, n)$ by invariant~\eqref{eq:inv-gcd}.
After such an unsynchronized reduction, the algorithm no longer modifies $u$: it halves $u$ only if $u$ is even, but $u$ is odd (the reductions happen only if $u$ and $v$ are both odd), and it subtracts $v$ from $u$ only if $v < u$, but $u = \gcd(u, v)$ never exceeds~$v$.
Since the algorithm updates $A$ and $B$ only when modifying $u$, they remain unchanged as well, which preserves invariant~\eqref{eq:inv-u} until the loop terminates.
In particular, the incorrect coefficients $C$ and $D$ are never used to update $A$ and $B$ and, thus, do not influence the result.
Our proof reflects this by a ghost variable tracking whether invariant~\eqref{eq:inv-v} still holds; once it does not, the loop invariant instead states that $u = \gcd(a, n)$ and that $u$ is odd.

\mypar{Fix}
\looseness=-1
We propose and implement a fix that synchronizes the reductions by using the same condition for both pairs of coefficients.
\Figref{go-sync-red-fix} shows the proposed fix for updating the coefficients~$A$ and $B$.
\code{x.add(y)} (respectively \code{x.sub(y)}) computes $x \mathrel{+}= y$ ($x \mathrel{-}= y$) on the arbitrary-length integer representation and return the carry (borrow) flag.
The branch condition checks whether the carry flag is set or $A \geq m$.
We prove (\cf \secref{verified-impl}) that updating $A$ and $B$ in this way is mathematically equivalent to \linerange{coef-AB-update-start}{coef-AB-update-end} in \algref{extgcd-fiatcrypto}.
This shows that our fix aligns Go's coefficient updates with those of \boringssl and \fiatcrypto at the algorithmic level, even though the latter two implement these updates using constant-time operations.

\begin{figure}[t]
\begin{gobracode}
carry := A.add(C)
B.add(D)
if choice(carry) == yes || A.cmpGeq(m) == yes {
  A.sub(m)
  B.sub(a)
}
\end{gobracode}
\caption{Proposed synchronized reduction for updating the coefficients~$A$ and $B$. Analogously, we propose the same fix for updating the coefficients~$C$ and $D$.}
\label{fig:go-sync-red-fix}
\end{figure}

\subsection{Input Domain Deviation}
\label{sec:spec-mismatch}
As the second~major deviation, we found that the Go implementation permits a larger input domain than \boringssl and the proven \fiatcrypto implementation.
This is critical as this larger input domain breaks the invariants on which the proof relies.

\Algref{extgcd-fiatcrypto} specifies as a precondition that $a < n$ must hold\footnote{Both \boringssl and \fiatcrypto state this precondition. In addition, \boringssl checks this condition at runtime and returns an error if it is not met.}.
The Go implementation, however, does not specify this requirement for its \gogcd function, which returns the \ac{GCD} value.
Dropping this requirement is natural as the mathematical definition of \ac{GCD} is symmetric in its two~arguments, and thus, does not require any ordering assumption.

Addressing this deviation requires proof changes only, not code changes.
We ran targeted tests to gain confidence that \gogcd produces correct results even when $a \geq n$, and we extended our formal proof to also cover the case $a \geq n$ by relaxing the invariants in the Go implementation.
Specifically, the invariants~\eqref{eq:inv-u}, \eqref{eq:inv-v}, $a < n$, and the bounds for the coefficients~$A$, $B$, $C$, and $D$ hold only when $a < n$.
In addition, we adapted the postcondition of \goextgcd to reflect the different guarantees that hold in the two~cases $a < n$ and $a \geq n$.
Only the former~case guarantees that the coefficient~$A$ is the modular inverse of $a$ modulo~$n$, while both~cases guarantee that $u = \gcd(a, n)$ holds.

\section{Verified Go Implementation}
\label{sec:verified-impl}
\looseness=-1
We verify correctness and termination of the existing and fixed extended \ac{GCD} implementations in Go but focus on the fixed implementation in this section.
All code, proofs, and continuous integration scripts are available open-source~\cite{extended-gcd-repo}.
First, we introduce the \gobra program verifier and illustrate necessary background on excerpts from our verified implementation in \secref{gobra}.
\Secref{verified-impl-spec} presents the specifications capturing correctness and termination, against which we verify the implementation using \gobra.
We cover our assumptions in \secref{verified-impl-assumptions}, evaluate our proposed code changes and verification effort in \secref{verified-impl-eval}, and discuss our results and the use of AI agents in \secref{verified-impl-discussion}.

\subsection{\gobra}
\label{sec:gobra}
\gobra~\cite{DBLP:conf/cav/WolfACOPM21} is a deductive, separation-logic-based program verifier for Go, which has been successfully applied to verifying parts of the WireGuard VPN protocol implementation in Go~\cite{DBLP:conf/ccs/ArquintSM023,DBLP:conf/sp/ArquintWLSSWBM23}, an Internet router for the SCION Internet architecture~\cite{DBLP:conf/ccs/PereiraKGLSWE0B25}, and a fork of the AWS SSM Agent~\cite{DBLP:conf/sp/ArquintKKDKM26}.
It translates annotated Go programs into the \viper intermediate verification language~\cite{DBLP:conf/vmcai/0001SS16}, which uses symbolic execution and discharges proof obligations via the \zthree SMT solver.

\mypar{Modular Verification}
\gobra performs \emph{modular} verification: each function is verified in isolation against its specification, which consists of preconditions (\code{requires}) and postconditions (\code{ensures}).
A function verifies successfully if executing the function body in an arbitrary state satisfying the precondition is guaranteed to result in a state satisfying the postcondition.
A caller must establish the preconditions of its callee and can safely assume the callee's postconditions.
This decomposition allows proofs to scale, since the verification of each function depends only on the specifications of its callees, not their implementations.

\mypar{Permissions}
\looseness=-1
\gobra uses separation logic~\cite{DBLP:conf/lics/Reynolds02} to reason about heap-allocated objects, side effects, and concurrency.
Every heap access requires that the current thread possesses the \emph{permission} for that heap object.
A permission for the heap object pointed to by pointer \code{ptr} is expressed as \code{acc(ptr, p)}, where \code{p} is a fractional permission~\cite{DBLP:conf/sas/Boyland03} in the interval $(0, 1]$.
A full permission ($p = 1$) grants both read \emph{and} write access, while any fraction strictly greater than zero and smaller than $1$ grants read-only access.
Permission for a particular heap object can be split and combined, \eg to share read access among multiple threads.
Since permissions are non-duplicable, a proof in separation logic guarantees memory safety and absence of data races: if two~threads have sufficient permissions to access the same heap object, they both must have a non-zero permission for this heap object, which is possible as long as the sum of these permissions does not exceed~$1$.
This ensures that these threads perform only read but no write accesses, unless the threads synchronize their accesses by using a suitable concurrency primitive\footnote{%
Proof-theoretically, a concurrency primitive can be used to allow threads to exchange permissions with each other, and, thus, obtain full permissions within a mutually-exclusive, critical section, which justifies certain write accesses therein.}.

\ifthenelse{\boolean{doublecolumn}}{%
\begin{figure}[t]
\begin{gobracode}
type Nat struct {
  limbs []uint
}
/*@
pred (n *Nat) Inv() {
  acc(n) && $\label{line:inv-acc-n}$
  (forall j int :: $\mathtt{0} \leq \mathtt{j} <{}$len(n.limbs) ==>$\label{line:inv-acc-limbs-start}$
    acc(&n.limbs[j])) && $\label{line:inv-acc-limbs-end}$
  let allLimbs := n.limbs[:cap(n.limbs)] in $\label{line:inv-cap-zero-start}$
  (forall j int :: len(n.limbs)${}\leq \mathtt{j} <{}$len(allLimbs) ==>
    acc(&allLimbs[j]) && allLimbs[j] == 0) $\label{line:inv-cap-zero-end}$
}
@*/
\end{gobracode}
\caption{Definition of the \code{Nat} struct that the Go standard library uses to represent arbitrary-length natural numbers and our invariant predicate.}
\label{fig:nat-inv}
\end{figure}
}{%
\input{sections/04_verified_implementation_nat_definition_singlecolumn}
}

\mypar{Predicates}
Permissions can be grouped into \emph{predicates}, such as the \code{Nat.Inv()} predicate~(\cf \figref{nat-inv}) used throughout our verification.
\code{Nat} is a struct type representing arbitrary-length natural numbers.
Internally, this type uses a slice of machine words, called \emph{limbs}, to represent the number in a little-endian manner.
A slice in Go provides a dynamic view into an underlying fixed-size array.
The \code{Nat.Inv()} predicate bundles the permissions for a \code{Nat}'s struct field~(\lineref{inv-acc-n}), all its limbs up to the slice's length~(\ifthenelse{\boolean{doublecolumn}}{\linerange{inv-acc-limbs-start}{inv-acc-limbs-end}}{\lineref{inv-acc-limbs}}), and the limbs between the slice's length and capacity~(\linerange{inv-cap-zero-start}{inv-cap-zero-end})\footnote{We split the permissions to the elements of the underlying array into two~separate forall quantified assertions for automation reasons.}.
In addition, \linerange{inv-cap-zero-start}{inv-cap-zero-end} state the invariant that limbs beyond the current length are zero, which was crucial for the correctness of the \code{Nat.reset} function until recently (\cf \secref{verified-impl-discussion}).

Callers pass permissions to \code{Nat.Inv()} in preconditions, \eg \code{n.Inv()} to grant the callee read and write access to \code{n}.
Passing a $p$~fraction of this invariant to a callee (\code{acc(n.Inv(), p)}) conceptually passes $p$~fractions of all permissions bundled in \code{Nat.Inv()}.
Predicates are opaque, meaning that accessing the permissions bundled in a predicate instance requires unfolding the predicate first, which exchanges the predicate instance for its definition.
To simplify the presentation, we use full permissions in the remainder of this paper.
However, we use more permissive specifications for our verified functions in the codebase, which distinguish between read-only and write access for inputs.

\begin{figure}[t]
\begin{gobracode}
/*@
ghost
opaque
requires n.Inv()
ensures  0 <= res && res < n.ValCount()
decreases
pure func (n *Nat) Repr() (res integer) {
  return ...
}

ghost
opaque
requires n.Inv()
ensures  0 <= res
decreases
pure func (n *Nat) AnnLen() (res int) {
  return unfolding n.Inv() in len(n.limbs)
}
@*/
\end{gobracode}
\caption{Ghost functions providing an abstraction for \code{Nat} by mapping it to its unbounded, numeric value and exposing its limbs' length. \code{Nat.Repr}'s postcondition states that $\texttt{res} \in [0,2^{\texttt{bits.UintSize} * \texttt{len(n.limbs)}})$.
}
\label{fig:nat-repr}
\end{figure}

\mypar{Ghost Code}
\looseness=-1
\gobra supports \emph{ghost code}: annotations that exist solely for verification and do not alter a program's runtime behavior.
Ghost code is delimited by \code{//@} (single-line) or \code{/*@ ... @*/} (multi-line) comments, making it invisible to the Go compiler.
Ghost annotations include ghost parameters, ghost variables, and ghost functions (marked \code{ghost}).
For instance, \code{Nat.Repr()} in \figref{nat-repr} is a pure function that returns the numeric value represented by a \code{Nat}'s limbs.
\code{pure} indicates that a function is free of side effects, which allows us to call it in specifications and loop invariants.

Since the length of the limbs is crucial, as manifested by numerous natural-language specifications referring to a \code{Nat}'s ``announced length'', we additionally define a ghost function~\code{Nat.AnnLen()}, which allows us to easily refer to the limbs' length in specifications and proofs.
To optimize verification time, we mark some functions as \code{opaque}, hiding their implementation by default. Their implementation is revealed at specific proof points only, via a \codebw{reveal} operation, to connect a \code{Nat}'s abstraction to its implementation. 

Ghost parameters are useful for avoiding existential quantification in specifications.
For example, \goextgcd (\cf \secref{verified-impl-spec}) returns the ghost value \code{BRepr} alongside its regular results, allowing the postcondition to state the B\'ezout identity
\texttt{u.Repr() == A.Repr() * a.Repr() - BRepr * m.Repr()}
without having to existentially quantify~\code{BRepr}.

\mypar{The \code{trusted} Keyword}
Functions annotated with \code{trusted} are assumed to satisfy their specification without verification.
We use \code{trusted} for bit-level operations on \code{Nat}, whose correctness depends on low-level arithmetic properties that \gobra cannot yet verify (\eg \code{cmpGeq}, \code{add}, \code{sub}, and \code{rshift1}).
In contrast, all our lemmata are either verified by \gobra or backed by \lean proofs included as comments in the source code.
The latter proved useful for non-linear arithmetic lemmata that \zthree could not handle.

To focus our verification efforts on \goextgcd and its clients, we marked several non-ghost functions in the \texttt{crypto\slash{}internal\slash{}fips140\slash{}bigmod} package as \code{trusted}.
We equip them with a trivially unsatisfiable precondition (\ie \code{requires false}) to make it explicit that these functions are irrelevant for \goextgcd \emph{and} their behavior is not yet faithfully captured in their specification.
Since this precondition can never be established, we ensure that these functions are not called from any verified code, preventing unsound assumptions about their behavior from creeping into the proof of \goextgcd.

\subsection{Specifications and Proofs}
\label{sec:verified-impl-spec}
Thanks to the code and proof fixes described in \secref{major-deviations}, we successfully verify the extended \ac{GCD} implementation in Go.
In this subsection, we present the specifications of the internal \goextgcd function and the two~functions \gomodinv and \gogcd, which are the two~clients of \goextgcd.
We end with covering key lemmata that we use throughout the proof.

\mypar{\goextgcd}
\Figref{extendedgcd-spec} shows the specification of \goextgcd against which we successfully verify the implementation using \gobra.
Our specification is a direct formalization of the natural-language specification in the standard library, which is included at the top of the code listing.
Due to our finding in \secref{spec-mismatch}, \linerange{extendedgcd-spec-inverse-start}{extendedgcd-spec-inverse-end} state that $u = A*a - B*n$ holds only if $a < n$.
Additionally, \code{decreases} instructs \gobra to also prove that the function terminates.

\ifthenelse{\boolean{doublecolumn}}{%
\begin{figure}
\begin{gobracode}
// extendedGCD computes u and A such that
// u = GCD(a, n) = A*a - B*n. u will have the size of
// the larger of a and n, and A will have the size of
// n. It is an error if either a or n is zero, or if
// they are both even.
//@ requires a.Inv() && n.Inv()
//@ ensures  a.Inv() && n.Inv()
//@ ensures  err == nil ==> u.Inv() && A.Inv()
//@ ensures  err == nil ==>
//@   u.Repr() == gcd(a.Repr(), n.Repr())
//@ ensures  err == nil && a.Repr() < n.Repr() ==>$\label{line:extendedgcd-spec-inverse-start}$
//@   u.Repr() == A.Repr()*a.Repr() - BRepr*n.Repr()$\label{line:extendedgcd-spec-inverse-end}$
//@ ensures  err == nil ==>
//@   u.AnnLen() == gmax(a.AnnLen(), n.AnnLen())
//@ ensures  err == nil ==> A.AnnLen() == n.AnnLen()
//@ decreases
func extendedGCD(a, n *Nat
) (u, A *Nat, err error /*@, ghost BRepr uint @*/)
\end{gobracode}
\caption{Specification for the verified \goextgcd function. For presentation purposes, we use full instead of fractional permissions for \code{a} and \code{n} and omit that \code{a} and \code{n} remain unmodified. \code{gmax} returns the maximum of its arguments.}
\label{fig:extendedgcd-spec}
\end{figure}
}{%
\input{sections/04_verified_implementation_extendedgcd_singlecolumn}
}

Within the implementation of \goextgcd, we prove its loop using the adapted loop invariants as described in \secref{spec-mismatch}.
For automation reasons, we hide the invariants~\eqref{eq:inv-u} and \eqref{eq:inv-v} (that hold in the case that $a < n$), in opaque pure functions and reveal their definitions only where necessary.
This does not impact soundness but speeds up verification significantly by avoiding exposing non-linear arithmetic to the SMT solver for most of the proof.
We use \texttt{u.Repr() + v.Repr()} as a termination measure for the loop, which instructs \gobra to prove that this expression strictly decreases in each iteration and is bounded from below, which implies that the loop terminates.

\mypar{\gomodinv and \gogcd}
\gomodinv computes the modular inverse of \code{a} modulo \code{n} and \gogcd computes the \ac{GCD} of \code{a} and \code{b}.
Both functions internally invoke the \goextgcd function, and we prove them correct against the specifications in \figref{inversevartime-gcdvartime-spec}.
Both functions store the result in \code{x}, which is also returned as the first return parameter~\code{r}.

\ifthenelse{\boolean{doublecolumn}}{%
\begin{figure*}
\begin{gobracode}
// InverseVarTime calculates x = a$\textcolor{\ghostcodebgcolor}{^{-1}}$ mod n and returns (x, true) if a is invertible. Otherwise, InverseVarTime
// returns (x, false) and x is not modified. a must be reduced modulo n, but doesn't need to have the same size. The
// output will be resized to the size of n and overwritten.
//@ requires x.Inv() && a.Inv() && n.Inv()
//@ requires a.Repr() < n.Repr()$\label{line:inversevartime-spec-a-reduced}$
//@ ensures  x.Inv() && a.Inv() && n.Inv()
//@ ensures  r == x
//@ ensures  ok ==> gcd(a.Repr(), n.Repr()) == 1
//@ ensures  ok ==> gcd(a.Repr(), n.Repr()) == x.Repr() * a.Repr() - BRepr * n.Repr()
//@ ensures  ok ==> x.AnnLen() == n.AnnLen()
//@ ensures !ok ==> x.Repr() == old(x.Repr()) && x.AnnLen() == old(x.AnnLen())$\label{line:inversevartime-spec-unmodified}$
//@ decreases
func (x *Nat) InverseVarTime(a *Nat, n *Modulus) (r *Nat, ok bool /*@, ghost BRepr uint @*/)

// GCDVarTime calculates x = GCD(a, b) where at least one of a or b is odd, and both are non-zero. If GCDVarTime
// returns an error, x is not modified. The output will be resized to the size of the larger of a and b.
//@ requires x.Inv() && a.Inv() && b.Inv()
//@ requires a.Repr() % 2 == 1 || b.Repr() % 2 == 1$\label{line:gcdvartime-spec-at-least-one-odd}$
//@ requires a.Repr() != 0 && b.Repr() != 0
//@ ensures  x.Inv() && a.Inv() && b.Inv()
//@ ensures  err == nil ==> r == x
//@ ensures  err == nil ==> x.Repr() == gcd(a.Repr(), b.Repr())
//@ ensures  err == nil ==> x.AnnLen() == gmax(a.AnnLen(), b.AnnLen())
//@ ensures  err != nil ==> x.Repr() == old(x.Repr()) && x.AnnLen() == old(x.AnnLen())
//@ decreases
func (x *Nat) GCDVarTime(a, b *Nat) (r *Nat, err error)
\end{gobracode}
\caption{Specifications for the verified \gomodinv and \gogcd functions.
We make similar presentation choices as in \figref{extendedgcd-spec}, and \code{Modulus.Inv()} is a predicate similar to \code{Nat.Inv()} that bundles permissions for a \code{Modulus} struct.
\code{old(e)} refers to the value of expression \code{e} in a function's pre-state, allowing us to express that a \code{Nat} is not modified in case of failure.
}
\label{fig:inversevartime-gcdvartime-spec}
\end{figure*}
}{%
\input{sections/04_verified_implementation_inversevartime_singlecolumn}}

\lineref{inversevartime-spec-a-reduced} directly encodes the standard library's natural-language requirement that \code{a} be reduced modulo \code{n}.
Similarly, \lineref{inversevartime-spec-unmodified} states that \code{x} is not modified on failure.
\gogcd follows the same pattern, where \lineref{gcdvartime-spec-at-least-one-odd} expresses that at least one of \code{a} or \code{b} must be odd, while the next line states that both \code{a} and \code{b} must be non-zero.

\mypar{Key Lemmata}
We define and prove eight~lemmata, which are crucial for the proof of \goextgcd and most of which involve non-linear integer arithmetic.
\gobra verifies four of them, while we provide machine-checked \lean proofs for the other four~lemmata.
For the latter four~lemmata, we include the \lean proof as comment in the codebase while marking the corresponding lemma as \code{trusted} for \gobra.
Two of these four~\lean-checked lemmata have direct counterparts in \fiatcrypto, while the other two state basic properties of modular arithmetic that \zthree cannot derive on its own.
Namely, 
\begin{align*}
    0 < b < a &\implies a \bmod b = (a - b) \bmod b \qquad\text{ and}\\
    0 \leq a \land 0 \leq b &\implies (a * b) \bmod 2 = (a \bmod 2) * (b \bmod 2) \text{.}
\end{align*}

\subsection{Assumptions}
\label{sec:verified-impl-assumptions}
Our verification rests on the following assumptions.

\mypar{\gobra and \zthree}
We trust the correctness of the \gobra verifier and the underlying SMT solver.
Since \gobra translates Go programs to \viper, which encodes verification conditions in SMT, a bug in any of these components could lead to unsound verification.
This risk is mitigated by choosing a widely used SMT solver, namely \zthree, and having substantial testing frameworks for both \gobra and \viper.

\mypar{Trusted Functions}
\looseness=-1
We assume that all functions annotated with \code{trusted} (and without \code{requires false}) satisfy their specifications.
These functions primarily perform bit-level operations.
Verifying their implementation would require better support for bit-level reasoning in \gobra, which is a promising direction for future work.

While we assume that these functions satisfy their specifications, these functions are relatively straightforward and consist of at most 13~lines of code each, making it feasible to review them manually and gain confidence in their correctness.
Similarly, our four~trusted lemmata (\cf \secref{verified-impl-spec}) are formalized in \lean.
While the \lean proofs are machine-checked and the translation from \gobra to \lean is straightforward and has been manually reviewed, there remains a small trust gap as we switch from one formalism to another.

\mypar{Overflows}
\looseness=-1
\gobra currently treats \code{int} and \code{uint} as unbounded types.
This means that our verification does not detect potential integer overflows in intermediate computations.
In our setting, this is mitigated by the fact that the \code{Nat} type manages multi-precision arithmetic internally.
In particular, the trusted functions are assumed to handle limb-level overflows and return carry and borrow flags accordingly.
Given these trusted functions, we prove that the verified functions correctly consider the returned carry and borrow flags for their computations.
Ongoing work in \gobra aims to improve the support for overflow checking, which would allow us to verify the absence of overflow bugs in the future.

\subsection{Evaluation}
\label{sec:verified-impl-eval}
We evaluate our work by measuring the performance impact of our proposed code change and the verification effort.

\mypar{Performance Evaluation}
\Secref{wrapping-bug} described a deviation of the existing Go implementation from the algorithm used in \boringssl and \fiatcrypto, which breaks the invariants in the \fiatcrypto proof.
We address this deviation with a fix (\cf \figref{go-sync-red-fix}), raising the question to what extent this fix impacts the implementation's runtime performance.
We answer this question by benchmarking the existing and fixed implementations of \goextgcd, since our proposed fix is limited to this function.
Since its clients \gomodinv and \gogcd remain unchanged and only handle errors and copy the result, we expect them to perform like \goextgcd.

\newcommand*{\runBenchPairs}{4}
\newcommand*{\runBenchCount}{10}
We use Go's built-in benchmarking framework to measure the runtime of \goextgcd for different limb sizes, namely $2^k$ for $k \in \interval[scaled]{0}{7}$.
For this purpose, we generate \num{\runBenchPairs}~input pairs~$(a, n)$ per limb size, where each limb in $a$ and $n$ is initially randomly sampled, and we ensure that the following properties hold:
$n$'s most significant limb is non-zero (otherwise, we resample this limb), $n$ is odd by setting the least significant bit to 1, and $0 < a < n$ by taking $a$ modulo $n$ and checking that the result is strictly positive (otherwise, we resample $a$ and start over).
For each input pair, we perform \num{\runBenchCount}~runs for both the existing and fixed implementations, where each run uses \code{testing.B.Loop}~\cite{go-loop-blog,go-loop-doc}, Go's preferred way of writing benchmarks.
\code{testing.B.Loop} repeatedly invokes \goextgcd with the same input pair until the total runtime exceeds the threshold of \qty{1}{\second} to amortize measurement overhead.

\begin{figure}[t]
\ifthenelse{\boolean{doublecolumn}}{}{%
    \begin{minipage}{.62\textwidth}
}
\includegraphics[width=\linewidth]{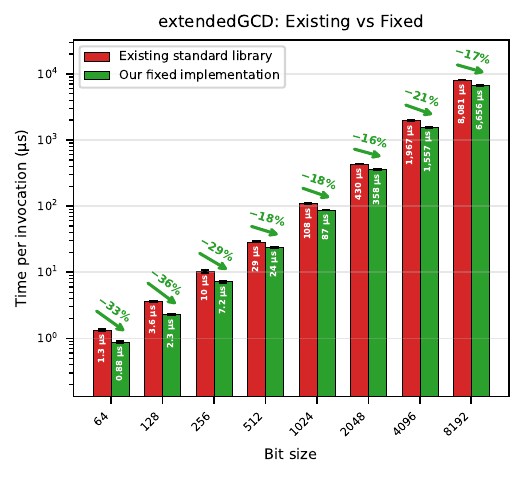}
\captionof{figure}{%
    Performance comparison between the existing and our fixed implementations for different limb sizes.
    Lower is better.
    Each bar shows the arithmetic average runtime across \num{\runBenchPairs}~input pairs and \num{\runBenchCount}~runs per pair, with the average labeled within each bar and error bars showing the standard deviation.
    The green arrows indicate the percentage speedup achieved by our implementation.
}
\label{fig:impl-performance}
\ifthenelse{\boolean{doublecolumn}}{}{%
    \end{minipage}
    \hfill%
\begin{minipage}{.35\textwidth}
    \centering
    \vspace{6.35em}
    \begin{tabular}{ l c c }
    \hline
    Function & \acsp{LOC} & \acsp{LOS}\\ 
	\hline
    \goextgcd & \num{68} & \num{203}\\
    \gomodinv & \num{10} & \num{18}\\
    \gogcd & \num{7} & \num{13}\\
    \code{syncAdd} & \num{8} & \num{63}\\
    Other Go functions & \num{69} & \num{84}\\
    Auxiliary definitions & -- & \num{131}\\
    Lemmata & -- & \num{466}\\
	\hline
    Total & \num{162} & \num{978}\\
    \hline
\end{tabular}

    \captionof{table}{%
        Overview of the verified \acptitle{LOC} and \acptitle{LOS} (incl.\ ghost code) in \gobra.
    }
    \label{tab:verified-code}
\end{minipage}
}
\end{figure}

\Figref{impl-performance} visualizes our results, showing the arithmetic average runtime and standard deviation for each limb size and implementation.
In particular, the standard deviation includes both the variability across \num{\runBenchPairs}~input pairs and the variability across different runs for the same input pair.
For all measured limb sizes, our fixed implementation is at least \qty{15.6}{\percent} faster than the existing implementation.
Across all benchmarks, we obtain a statistically significant, geometric mean performance improvement of \qty{23.98}{\percent}.

The main difference between the existing and fixed implementations beyond the synchronization of the coefficient updates (\cf \secref{wrapping-bug}) is the way intermediate results are stored~(1) and computed~(2).
In terms of storage~(1), each loop iteration that triggers the loop body's first~part (\linerange{both-odd-start}{both-odd-end} in \algref{extgcd-fiatcrypto}), allocates in the existing implementation two~buffers for storing intermediate results if the intermediate results use up to \num{2048}~bits, and four~buffers otherwise.
In the latter case, the number of allocations doubles because Go's \code{Nat} type allocates a \qty{2048}{\bit} buffer in its constructor, which ends up being too small for storing the intermediate results, leading to an additional buffer allocation per intermediate result.
In contrast, our fixed implementation performs in-place updates and, thus, allocates zero~buffers in the loop body.
In terms of computations~(2), the existing implementation performs, in addition to the operations given on \linerange{both-odd-start}{both-odd-end}, two~zeroing operations for limbs sizes $\leq 2048$~bits, two~copying operations, and two~conditional copy operations, each iterating over all limbs of a \code{Nat}, while saving one comparison operation.
In contrast, our fixed implementation performs only the operations as stated on \linerange{both-odd-start}{both-odd-end} in \algref{extgcd-fiatcrypto}.

\mypar{Proof Evaluation}
We successfully verify the \goextgcd implementation and its callers \gomodinv and \gogcd against their specifications using \gobra.
Our verified code contains both variants of the coefficient update and selects one of them via a global boolean variable.
Since this variable is left unspecified, the proof covers both the existing and fixed implementations.
In order to prove termination and their correctness, we require specifications, which include loop invariants, define and prove lemmata (\cf \secref{verified-impl-spec}), and state and assume specifications for trusted functions (\cf \secref{verified-impl-assumptions}).
Here, we report on the proof size and the verification time.

\ifthenelse{\boolean{doublecolumn}}{%
\begin{table}[t]
    \centering
    
    \caption{%
        Overview of the verified \acptitle{LOC} and \acptitle{LOS} (incl.\ ghost code) in \gobra.
    }
    \label{tab:verified-code}
\end{table}
}{}

\looseness=-1
\Tabref{verified-code} summarizes our verified code, measured in \acp{LOC} and \acp{LOS}, where the latter includes ghost code.
A line is counted as a line of code \emph{and} a line of specification if it contains both code and specification elements, which is, \eg the case for a function signature that includes a ghost parameter, like the last line in \figref{extendedgcd-spec}.
\code{syncAdd} implements the coefficient updates using synchronized reduction, whose implementation without any proof-related specifications is shown in \figref{go-sync-red-fix}.
Auxiliary definitions and lemmata are exclusively ghost code.
In particular, auxiliary definitions consist of predicates and pure functions (\cf \secref{gobra}), while lemmata derive, \eg mathematical properties about \ac{GCD} or non-linear arithmetic that are crucial for the proof of \goextgcd.
While \num{22}~lines of lemmata are trusted, they are backed by \num{36}~lines of \lean proofs establishing their correctness.
Since \tabref{verified-code} reports \emph{verified} lines, the table does \emph{not} include \code{trusted} functions that are neither verified by \gobra nor backed by \lean proofs, which include bit-level operations on \code{Nat}.
We have 14~trusted functions like \code{Equal}, \code{IsZero}, \code{IsOdd}, and \code{add}, whose specifications and implementations are relatively straightforward and easy to review manually, amounting to a total of \num{117}~\acp{LOC} and \num{98}~\acp{LOS}.

All functions, auxiliary definitions, and lemmata together are verified in \qty{33.8}{\second}.
We have measured the verification times by computing the \qty{10}{\percent} Winsorized mean of the wall-clock runtime across \qty{20}{verification} runs on a 2024~Apple MacBook Pro with M4~Pro processor, macOS Golden Gate~27.0, \gobra~commit~eff4f89, and \zthree~4.8.7~arm64.
This short verification time allows us to perform continuous verification, \ie we integrated \gobra into a GitHub continuous integration workflow to verify the implementation after each code change~\cite{extended-gcd-ci}.

\subsection{Discussion}
\label{sec:verified-impl-discussion}
Modern program verifiers like \gobra are powerful tools that significantly increase the confidence in the correctness of complex implementations, such as the extended \ac{GCD} algorithm in Go's standard library.
Although closing the remaining trust gap would require better support for bit-level operations in \gobra, we verified the algorithm's implementation by building on a set of small trusted functions.
Effectively, we break down the correctness of the overall implementation, whose correctness argument goes beyond what we can realistically expect from code reviewers, into the correctness of a handful of trusted functions.
We are convinced that these trusted functions are small and straightforward enough, such that reviewers can convince themselves of their correctness \wrt their specifications, which state the expected behavior in mathematical terms.
This achieves a good balance between verification effort and the gained confidence in the overall implementation's correctness.

Besides providing a means to break down the overall correctness argument, program verifiers impose a discipline of reasoning about correctness that helps uncover subtle issues.
We initially worked on the go1.24.0 version of the codebase, and realized that the implementation's correctness relies on an implicit invariant about the \code{Nat} type, namely that all limbs beyond the announced length must be zero; otherwise, changing the announced length could lead to a change in the represented value.
Our initial predicate definition was too weak to capture this invariant, which led to the addition of \code{allLimbs[j] == 0} on \lineref{inv-cap-zero-end} in \figref{nat-inv}.
While the Go developers independently discovered and addressed this brittleness~\cite{go-extended-gcd-inv-bug} within about three~months, the discipline of writing down invariants and reasoning about how the implementation relates to them led us to the same discovery in much less time.

\mypar{Experience with Agentic Verification}
We used Anthropic's AI agent Claude Code throughout the verification effort, which reduced the manual effort significantly.
We estimate that we spent around two~person-weeks on the verification, which includes prompting the agent.
We make the following observations.

Applying the agent was effective because \gobra produces insightful error messages that the agent could interpret and act upon.
This allowed the agent to autonomously iterate on both syntactic (\eg missing ghost arguments) and semantic issues, where the latter includes too weak or too strong preconditions, postconditions, and loop invariants as well as missing lemma invocations.
While the agent figured out on its own that moving some non-linear arithmetic reasoning into lemmata helps in avoiding timeouts in \zthree, we taught the agent about \code{opaque} \code{pure} functions to further reduce the impact of non-linear arithmetic, after which the agent effectively used them for the loop invariants~\eqref{eq:inv-u} and \eqref{eq:inv-v}.

When the agent repeatedly failed to prove the invariants for the existing coefficient update, it correctly suggested that the independent reduction of the coefficients might be the root cause.
This hint pointed us in the right direction to identify the unsynchronized subtractions deviation~(\secref{wrapping-bug}).
However, the agent initially treated the implementation as authoritative---assuming the code was correct and looking for a matching proof---rather than questioning the implementation's correctness.
We had to prompt the agent several times to compare the Go implementation with \fiatcrypto's algorithm and proof before it eventually identified the deviation.
Once identified, the agent was able to suggest a fix and successfully verify the fixed implementation.
We completed the proof of the fixed implementation first.
When we asked the agent in April~2026 to adapt this proof to the existing implementation, it failed.
When we repeated this attempt in August~2026 with a newer version of the agent, it succeeded.

While the agent eventually produced a valid proof, we performed several iterations of manual cleanup to combine, simplify, and rename the generated lemmata.
Hence, the final proof structure that uses two~invocations of an \code{opaque} \code{pure} function to express the invariants~\eqref{eq:inv-u} and \eqref{eq:inv-v}, and a few high-level lemmata is significantly cleaner than the intermediate versions that the agent produced.
Understanding the implementation and the generated proof as well as performing this cleanup constituted the majority of the two person-weeks we spent on the verification effort.
Nevertheless, we estimate that the agent significantly reduced the verification effort by providing a valid proof that we could then clean up, rather than requiring us to develop the proof details from scratch.

\section{Related Work}
\label{sec:related-work}
We compare our work to related efforts in the area of verified cryptographic implementations, with a focus on the extended \ac{GCD} algorithm and modular inverse computations.

\fiatcrypto~\cite{DBLP:conf/sp/ErbsenPGSC19} is closest to our work, as it also provides a verified extended \ac{GCD} implementation.
\fiatcrypto is a framework for the \rocq proof assistant that allows one to express and prove algorithms in a high-level language and then extract efficient C~code.
While \boringssl has adopted Curve25519 and P-256 implementations generated by \fiatcrypto and, thus, benefits from \fiatcrypto's guarantees, the story for the extended \ac{GCD} implementation is different.
\boringssl uses a handwritten C implementation of the extended \ac{GCD} algorithm, which has been ported and proven correct in \fiatcrypto.
Since there is no formal link between \boringssl's and \fiatcrypto's extended \ac{GCD} implementations, subtle differences or regressions may go unnoticed.
In contrast, we reason about implementations at the code level, which allows us to verify \emph{existing} implementations and does not require implementers to replace their handwritten implementations with generated ones.

\evercrypt~\cite{DBLP:conf/sp/ProtzenkoPFHPBB20} is a verified cryptographic library that bundles \haclstar~\cite{DBLP:conf/ccs/ZinzindohoueBPB17} and \vale~\cite{DBLP:conf/uss/BondHKLLPRST17}.
\haclstar is a library written and verified in \fstar~\cite{DBLP:conf/popl/SwamyHKRDFBFSKZ16} and extracted to C, while \vale is a tool for writing low-level code in a C-like language; this code is verified in either \dafny~\cite{DBLP:conf/lpar/Leino10} or \fstar and extracted to assembly.
While we are not aware of any \ac{GCD} or modular inverse implementations in \vale, \haclstar provides only limited support for computing modular inverses, which is insufficient for RSA key generation.
In particular, \haclstar implements two~different approaches to compute the modular inverse, neither of which is as general as the extended \ac{GCD} algorithm (which we verify).
\code{mod_inv_limb} computes the modular inverse but is limited to single-limb moduli (we support multi-limb moduli), while \code{bn_mod_inv_prime} computes the modular inverse for multi-limb numbers using Fermat's little theorem and, thus, requires a prime modulus (which we do not require and is not the case for RSA key generation).
The same limitations apply to \libcrux~\cite{libcrux}, a Rust cryptographic library, because the library provides Rust implementations of the same two~functions by translating their \haclstar implementations to Rust.

The Formosa Crypto project~\cite{formosa-crypto} comes with a language and a provably semantics-preserving compiler called \jasmin~\cite{DBLP:conf/ccs/AlmeidaBBBGLOPS17} to write fast and secure low-level assembly code.
To prove properties about \jasmin code, the compiler either checks certain properties itself, \eg using a type-system to enforce constant-time programming, or extracts an \easycrypt~\cite{DBLP:conf/crypto/BartheGHB11} model about which one can prove properties using the \easycrypt proof assistant.
While \jasmin's arbitrary-length number library \libjbn~\cite{cryptoeprint:2023/752} computes modular inverses using Fermat's little theorem (thus, the same limitations apply as for \haclstar), we are not aware of any executable \jasmin implementation of the extended \ac{GCD} algorithm.

\section{Conclusions}
\label{sec:conclusions}

We have applied \gobra to the \goextgcd implementation in Go's standard library (\texttt{crypto/internal/fips140/bigmod}), uncovering two~deviations that each break the invariants on which \fiatcrypto's proof relies.
Our suggested fixes resolve both deviations, while significantly improving the implementation's performance.
Using \gobra, we prove both the existing and the fixed implementations correct against its formal specification, which mirrors the natural-language specification provided in the Go standard library.
Extending \gobra with stronger support for bit-level reasoning would enable us to verify functions that are currently trusted, thereby reducing our trust assumptions and further increasing confidence in the implementation.

\ifnum\chosenlayout=\acmlayout\relax
\begin{acks}
We thank Filippo Valsorda for the helpful discussions.
This work has been supported by a Singapore Ministry of Education~(MoE) Tier 3 grant \quotes{Automated Program Repair}, MOE-MOET32021-0001.
The author declares a non-financial interest as a developer of \gobra.
\end{acks}
\fi

\ifnum\chosenlayout=\lncslayout\relax
\begin{credits}
\subsubsection{\ackname}
We thank Filippo Valsorda for the helpful discussions.
This work has been supported by a Singapore Ministry of Education~(MoE) Tier 3 grant \quotes{Automated Program Repair}, MOE-MOET32021-0001.

\subsubsection{\discintname}
The author is a developer of \gobra.
\end{credits}
\fi

\ifnum\chosenlayout=\acmlayout\relax
  \bibliographystyle{ACM-Reference-Format}
  \balance
\fi
\ifnum\chosenlayout=\lncslayout\relax
  \bibliographystyle{splncs04}
\fi
\bibliography{bibliography}

\end{document}